\documentclass[12pt]{article}
\input epsf.sty
\def\be{\begin{equation}} \def\ee{\end{equation}}
\def\bea{\begin{eqnarray}} \def\eea{\end{eqnarray}} \def\ba{\begin{array}}
\def\ea{\end{array}} \def\ben{\begin{enumerate}} \def\een{\end{enumerate}}

\newcommand{\eqn}[1]{(\ref{#1})}

\newcommand{\hepth}[1]{{\tt hep-th/{#1}}}

\def\l{\lambda}
\def\m{\mu}

\def\n{\nu}

\def\br{\nonumber\\}

\begin{document}
{}~
\hfill\vbox{\hbox{hep-th/yymm.nnnn} \hbox{\today}}\break

\vskip 3.5cm
\centerline{\large \bf
Galilean anti-de-Sitter spacetime in Romans theory}
\vskip .5cm

\vspace*{.5cm}

\centerline{  Harvendra Singh}

\vspace*{.25cm}
\centerline{ \it  Theory Group, Saha Institute of Nuclear Physics} 
\centerline{ \it  1/AF, Bidhannagar, Kolkata 700064, India}
\vspace*{.25cm}

\vspace*{.5cm}

\vskip.5cm
\centerline{E-mail: h.singh [AT] saha.ac.in }

\vskip1cm

\centerline{\bf Abstract} \bigskip

The Romans type IIA theory is the only known example of  10-dimensional 
maximal supergravity where (tensor) fields are explicitly massive.
We provide an example of a non-relativistic anti-de Sitter 
$NRadS_4\times S^6$ background as a solution in 
massive type IIA. A compactification of which on $S^6$ gives 
immediately the prototype NRadS background in $D=4$ which is proposed to 
be dual to `cold atoms' or unitary fermions on a wire.      

\vfill 
\eject

\baselineskip=16.2pt

\section{Introduction}
Recently, the applications of AdS/CFT holography [1,2,3] to strongly 
coupled 
condensed matter systems, which 
show scaling behaviour near quantum critical points, have taken a big leap
[4-15]. In studying the scaling behaviour near critical points one 
considers 
the nonrelativistic limit of AdS/CFT which exhibits a reduced conformal 
symmetry or `Schroedinger group' \cite{mehen}. Its main 
applications to 
study strongly coupled fermionic 
systems at finite density has been  called as `AdS/Atoms' cold and hot 
[4,5]. \footnote{ A closely related framework on Galilean symmetry was 
studied much earlier by \cite{horvathy}.}  For the study 
of finite 
temperature properties like phase transitions, transport and viscosity 
etc one, however, 
needs to include black holes in AdS backgrounds [6,7].\footnote{See for 
example \cite{herzogrev,denefrev} for a list of many previous works.} 
For superconductivity the dual
non-relativistic AdS geometry generally involves 
spontaneously broken Higgs  phases where the Abelian gauge field 
becomes massive. 

Recently, there 
have been 
many examples where  non-relativistic anti-de-Sitter (NRadS) geometries 
can be embedded 
in type IIB and 
M-theory, see for example \cite{marte, 
Pal,denefrev,donos,gaunt,gubser}. 
While there have been no 
attempts to our knowledge where the same has been worked out for 
 massive type IIA supergravity.     
The Romans' massive IIA theory is the only known example of a 
10-dimensional 
maximal supergravity where (tensor) fields are explicitly massive to 
begin with. 
Thus massive type IIA  provides an unique  case to look for a 
NRadS 
solutions and study  dual Galilean field theory.

In this short note we provide an example of 
$NRadS_4\times S^6$ background as a solution in 
massive type IIA supergravity. A compactification of which on $S^6$ gives 
immediately the prototype NRadS background in $D=4$ which is proposed to 
be dual to non-relativistic field theory of cold atoms.      
The paper is organised as follows. In the section-2 first we review the 
relevant 
aspects of massive type IIA  sugra action and then obtain the 
non-relativistic 
AdS solution. We also discuss its reduction to four dimensions in 
section-3.
The brief discussion is given in 
section-4.

\section{Galilean solution of Romans theory}

The massive type IIA sugra action \cite{roma} 
is given  by
\bea
S={1\over 2 k^2}\int d^{10}x \sqrt{-g}\bigg[ R- {1\over2} 
(\partial_\m\phi)^2 
-{e^{-\phi}\over 2} (H_{3})^2  
-{e^{3\phi/2}\over 2} (G_{2})^2  
-{e^{\phi/2}\over 2} (G_{4})^2  
-{e^{5\phi/2}\over 2} m^2 \bigg]
\eea
where we have left out the topological terms
as those are vanishing 
for the type of backgrounds we are going to study,  for details see 
\cite{berg,hs}. Various field strengths 
are
\bea
H_3=dB_2,~~~G_2=dC_1 +m B_2, ~~~G_4=dC_3 +B_2\wedge dC_1 + {m\over2} 
B_2\wedge B_2
\eea
where $m$ is the mass parameter in the theory. The 2-rank tensor field 
is explicitly massive with mass square $m^2$,  
and there is a potential term $\propto m^2$ for the dilaton 
field. The potential is  due to the requirement of maximal supersymmetry 
in the 
10-dimensional theory \cite{roma}. However, as soon as the mass vanishes 
the potential 
altogether vanishes and the theory reduces to ordinary type IIA 
supergravity in ten dimensions. 

The massive type IIA theory does not admit any Minkowaskian vacuum 
solution 
\cite{roma}.    
Instead it is known that theory admits $1/2$-supersymmetric domain-wall 
solutions,
also called D8-branes \cite{Pol1,berg}, and other supersymmetric flux 
vacua 
like $(D6,D8)$ and $(D4,D6,D8)$ bound states \cite{hs,hs1}. 
Particularly interesting solution for this paper are the Freund-Rubin 
solutions 
$AdS_4\times S^6$ which are supported by a constant 4-form 
flux \cite{roma}
\bea\label{sol1}
&&ds^2=L^2\left({-2dx^{+}dx^{-}+dy^2+dz^2\over
z^2} +{5\over 2} d\Omega_6^2\right), \br
&&\phi=\phi_0, ~~~~G_{+-yz}= c L^4 z^{-4}
\eea 
with $2L^{-2}=m^2 g_s^{5/2}$, $c^2=5 m^2 g_s^2$, and $g_s=e^{\phi_0}$ 
being the string coupling constant. The $AdS_4$ metric is given with 
light-cone coordinates while $d\Omega_6^2$ is
the metric of  unit  
six-sphere. This is an example of a 
non-supersymmetric solution.
 
\subsection{Massive string (dust)} 
Recently, there have been several examples of non-relativistic geometries 
[4,5] which are proposed to be dual to condensed matter 
phenomenon like unitary fermion systems, superfluidity and 
superconductivity etc. An observed  
common feature has been that in order to obtain non-relativistic solutions 
one needs to 
introduce some
 massive (vector) fields propagating in the geometry. Hence in massive 
type IIA theory we 
must consider a 2-rank 
tensor, the only massive field in the theory, to be nontrivial.
The above action in the tensorial notation is given by
\bea\label{eq2}
S={1\over 2 k^2}\int d^{10}x \sqrt{-g}\bigg[ R- {1\over2} 
(\partial_\m\phi)^2 
-{e^{-\phi}\over 2.3!} (H_{\m\n\l})^2  
-{e^{3\phi/2}\over 2.2!} (G_{\m\n})^2 \br  
-{e^{\phi/2}\over 2.4!} (G_{\m\n\l\sigma })^2  
-{e^{5\phi/2}\over 2} m^2 \bigg],
\eea
with 
\bea
&& H_{\m\n\l}=3\partial_{[\m} B_{\n\l ]},~~~ 
G_{\m\n}=2\partial_{[\m}C_{\n]} + m B_{\m\n} , \br  &&
G_{\m\n\l\sigma}=4\partial_{[\m} C_{\n\l\sigma ]}+ 6
B_{[\m\n}G_{\n\sigma ]} -
3{m}B_{[\m\n}B_{\n\sigma ]}.
\eea
To solve the equations of motion obtained from action \eqn{eq2} 
we make the following ansatz for $NRadS_4\times S^6$ 
geometry and fields
\bea\label{sol2}
&&ds^2=L^2\left(-{2\over z^{2a}} 
(dx^{+})^2+{-2dx^{+}dx^{-}+dy^2+dz^2\over
z^2}  +{5\over 2} d\Omega_6^2 \right) ,\br
&&\phi=\phi_0, ~~~G_{+-yz}= c L^4/z^4 \br
&& 
B_{+ y}=  f(z), ~~~C_{+}= g(z) \ ,
\eea 
where $f(z), ~g(z)$ are functions which are to be determined next. 
Note that with this choice of the tensor fields 
$SO(1,3)$ Lorentz invariance is explicitly broken. The background will 
involve D0, D8 branes with 4-form flux alongwith fundamental strings 
stretched along the $y$ direction.  (One may, however, 
consider $B_{+z}$ to be non-zero and set $C_{+}=0$ by using the 
Stueckelberg gauge invariance.)
Given the
choice of $B_{\m\n}$ and $C_{\m}$ as in \eqn{sol2}, one finds that the 
invariants 
$(G_{\m\n})^2$ and 
$(H_{\m\n\l})^2$ 
vanish identically. Due to this the dilaton equation of motion
remains unchanged and is simply written as
\be\label{dil1}
\nabla \phi - V'(\phi)=0
 \ee
with the dilaton potential as
\be 
V(\phi)=-
{c^2\over2}e^{\phi/2}+ 
{m^2\over2}e^{5\phi/2}. \ee
Hence a constant dilaton solution $\phi=\phi_0$ will still be fixed by 
the 4-form flux and the mass parameter
 as in the relativistic case \eqn{sol1}, that is  $c^2=5 m^2 
g_s^2$.  It means that by making above 
non-relativistic deformations, the vital parameter like string coupling 
constant $g_s$ are not affected. Notice that these parameters determine 
the overall curvature of the spacetime which we have to keep small 
irrespective of the non-relativistic limit (deformation). Next
the Einstein equations for the metric \eqn{sol2} along the 
six-sphere 
also remain unchanged and those are solved if we set
  $2L^{-2}=m^2 g_s^{5/2}$ as earlier.
The equation involving $R_{++}$ component however gets modified now 
because 
$T_{++} \propto  g_{++}$  receives contributions from the 
gauge potentials $B_2$ and $C_1$. 

The $B$-field equations are solved if we take
\be\label{eq1a}
g(z)= {\sqrt{5}m\over a} {q \over z^{a}},~~~
f(z)= {q \over z^{a+1}} \ .
\ee
for the exponents being $a=3$ and $a=-4$ only. The last remaining constant 
$q$ is determined by 
the equation involving 
$R_{++}$. We do find
 that the non-relativistic solutions 
\eqn{sol2} exist provided 
\bea\label{eq2a}
q=&& \pm \sqrt{2 g_s } L^2 , ~~~ {\rm for} ~~~a=3 \br
q=&& \pm 2\sqrt{7 g_s \over 3} L^2 , ~~~ {\rm for} ~~~a=-4 \ .
\eea
From this we clearly see that the nonrelativistic matter (dust) 
responsible for the AdS
 deformation is nothing but made up of the massive {\it strings} 
 stretched along  $y$ direction. 
As the boundary of  $NRadS_4$ space is located at $z=0$, near the boundary 
the 
metric and fields are divergent for $a=3$ case but everything is fine in 
the 
interior of the spacetime. While for $a=-4$ it is other way round.
Nevertheless the invariants appearing in the action 
\eqn{eq2} stay finite every where in both the cases. 

Although our solutions \eqn{sol2} have an scaling (dilatation) symmetry
\be
x^{+}\to \lambda^a x^{+},~~
x^{-}\to \lambda^{2-a} x^{-},~~
y\to \lambda y,~~
z\to \lambda z
\ee
with  exponents $a=3,~-4$.
But these solutions have no special conformal symmetry as that arises
only when the 
exponent is $a=2$ \cite{son, bala}. 

\section{Compactification}

The four-dimensional AdS supergravity can be obtained by compactification 
of the action \eqn{eq2} on a six-sphere.
On consistent truncation and keeping only the relevant spacetime 
components, it 
will give us  following 
$(3+1)$-dimensional effective action 
\bea\label{eq3}
S_{4}\sim &&\int d^{4}x \sqrt{-g}\bigg[ R 
-{1\over 2.3!}{1\over g_s}(H_{\m\n\l})^2  
-{1\over 2.2!} g_s^{3\over 2}(G_{\m\n})^2 +\Lambda 
\bigg] \br 
&&-{\sqrt{5}m\over 2!2!} g_s^{3\over2} \int 
d^4x\epsilon^{\m\n\l\rho}(
B_{\m\n} G_{\l\rho}- {m\over 2}
B_{\m\n} B_{\l\rho}) \ ,
\eea
where we have kept only the tensor field and the 1-form and have 
integrated out the 4-form
field strength which is a volume form over spacetime. We have normalised 
$\epsilon^{0123}=1$.
The four dimensional cosmological constant 
expressed in 10-dimensional variables is given by
$\Lambda= 3m^2{g_s^{5/2}}={6\over L^2} $. 
It is important to note 
that from 10-dimenisonal point of view, $m$ is  fixed by the 4-form 
flux as $c^2=5 m^2 g_s^2$. \footnote{ In other words, by the number of 
$N_c$ D2 and $N_f$ D8-branes (usually in type I' picture $(N_f\le 16)$ 
\cite{Pol2})} But the curvature of 
spacetime will remain small so long as 
string coupling remains weak $(g_s<1)$. 

As it is obvious that $G_{\m\n}$ is an auxiliary field srength. We first 
integrate it out using its field equation
\be
\star G_2+ \sqrt{5} m  B_2=0
\ee
where $\star$ is a Hodge-dual in four dimensions.
Upon integration the action \eqn{eq3} reduces to
\bea\label{eq4}
S_{4}\sim &&\int d^{4}x \sqrt{-g}\bigg[ R 
-{1\over 2.3!} {1\over g_s}(H_{\m\n\l})^2  
-{5m^2\over 2.2!} g_s^{3\over 2}( B_{\m\n})^2 +\Lambda 
\bigg] \br &&
+ {\sqrt{5}m^2\over 2 (2!)^2} g_s^{3\over2} \int 
d^4x\epsilon^{\m\n\l\rho}
B_{\m\n} B_{\l\rho} \ .
\eea
Let us write down the equation of motion of $B_2$ obtained from varying 
the action \eqn{eq4} 
\be\label{oi01}
 {1\over g_s}d\star H_3- m^2 g_s^{3\over2}(5\star B_2 - \sqrt{5} 
B_2)=0
\ee
In order to see that it gives the same 
effective action as studied by \cite{son,bala}
we will define a  massive gauge (Proca) field
through a duality relation in four-dimensions as
\be\label{oi0}
{1\over g_s}\star H_3=  dX + m_0 A_1,
\ee
where $X$ is the  axion field and parameter $m_0$ will get specified next. 
It  is 
the Goldstone mode introduced so that 
there is a shift (Stueckelberg) invariance
\be
 \delta X=-m_0\l,~~\delta A_1=d\l .
\ee With this gauge symmetry the axion can always be 
eaten up by the gauge field. The equations \eqn{oi0} and \eqn{oi01} imply
\bea\label{oi02}
&&F_2\equiv dA_1 = {m^2 g_s^{3\over2}\over m_0} ({5}\star B -\sqrt{5} 
B) .
\eea  
It leads to a Proca equation
\be\label{oi03}
d\star F_2+ 6 m^2 g_s^{5\over2} \star A=0
\ee
Thus the gauge field has the mass square $6m^2  
g_s^{5\over2}$. 
We can also get this equation from a simple
action 
\bea\label{eq3a}
S_4\sim\int d^{4}x \sqrt{-g}\bigg[ R 
-{1\over 2.2!} (F_{\m\n})^2  
-{g_s\over 2} (\partial_\m X+m_0 A_{\m})^2 +{6\over L^2} \bigg]
\eea
provided we identify $m_0^2 = 6 m^2 g_s^{3\over2}={12\over g_s L^2}$.
This kind of action has been the starting point of many 
non-relativistic AdS geometries holographically
dual to the strongly coupled condensed matter phenomena \cite{son,bala}. 
The crucial difference however is that the mass of the Proca field is 
fine tuned 
to the negative
cosmological constant in the action. This comes from the 
consistent embedding of 
the theory in the Romans supergravity in ten dimensions.

The 4-dimensional non-relativistic vacua of \eqn{eq3a}
can be written as 
\bea\label{sol21}
&&ds^2=L^2\left(-{2\over z^{2a}} 
(dx^{+})^2+{-2dx^{+}dx^{-}+dy^2+dz^2\over
z^2} \right)  ,\br
&& 
A_{+}= {q(a+1)\over \sqrt{12 g_s} L}{1\over z^a}, ~~~X=0 \br
\eea 
where $q$ is as in eq. \eqn{eq2a}.

\subsection{Hierarchy of scales}
While truncating to the action \eqn{eq3} we kept dilaton fixed to 
its vacuum 
value $\phi=\phi_0$. Now if we allow infinitesimal perturbations in 
the dilaton field around its
anti-de Sitter 
minima, say $\phi=\phi_0+\rho$, keeping everything else fixed, we  
find from \eqn{dil1} that the 
fluctuation $\rho$ has got positive mass square given by $5/L^2$. 
Including the 
fluctuating mode of dilaton we may also write to the leading order
an action
\bea
S_4\sim\int d^{4}x \sqrt{-g}\bigg[ R +{6\over L^2} 
-{1\over 2.2!} (F_{\m\n})^2  
-{1\over 2} (\partial_\m \chi +m_A A_{\m})^2 
-{1\over2}(\partial_\mu\rho)^2-{5\over  L^2}{\rho^2\over 2} 
\bigg]
\eea
where by simple scaling $\chi=\sqrt{g_s} X,~~m_A^2=g_s m_0^2$.
The above action is nothing but represents a $U(1)$ field coupled to a 
complex scalar field with real components $(\rho,\chi)$
in an spontaneously broken Higgs vacua. It is interesting in the 
sense that  broken phases like this represent  dual
superconductors \cite{Hartnoll:2008kx}. It is remarkable that it
could also
be realised in Romans' massive  theory.
We note doen the characteristic ratio
\be
{m_A^2\over m_\rho^2}={12\over 5}.
\ee
The hierarchy of mass scales goes as
\be
m_\rho^2>m_{BF}^2 \ .
\ee
The Breitenlohner-Freedman (BF) bound emphasizes that the 
mass square of a scalar in $AdS_{d+1}$ should be larger than $-{d^2\over 4 
L^2}$.  
Thus the mass hierarchy is  satisfactory from the AdS stability point 
of view.

\subsection{Scaling solution}
We now look for the scaling solutions near the boundary. The 
isometry direction $X^{-}$ can be taken periodic and the corresponding 
Kaluza-Klein  momentum, $P^{+}=M$, modes will generate an spectrum.  
In a given momentum sector $M$,  there
are no scaling solutions near the boundary for NRadS geometries if the 
exponents $a>2$ \cite{bala}. However in zero light-cone momentum sector
 we will still have a spectrum of  non-normalisable operators. We  find  
these operators with non-relativistic geometry exponent $a=3$. 
If the bulk gauge 
field $A_{\m}(x;z)$ behaves as $\sim z^{-\lambda_+}$ near the 
boundary, the 
corresponding conformal 
operator  is the conserved current
 $J^{\m}(x)$ in the boundary theory having conformal dimension 
$\Delta_{+}=d-1+\lambda_{+}$ \cite{Witten:1998qj}. We can easily read from 
our $NRadS_4$ solution 
$\lambda_{+}=a=3$. The 
operator dimension for a spin-1  field can also be obtained 
 as 
\be
\Delta_\pm={d\over 2}\pm\sqrt{ ({d-2\over 2})^2 + m^2 L^2}  
\ee
With mass as $m^2=12 L^{-2}$ and $d=3$, we get the conformal dimension 
of the current operator  to be $\Delta_{+}= 5$. Thus 
the inclusion of massive gauge fields in  Romans 
theory gives rise to an irrelevant non-relativistic deformation of the 
original  
$(2+1)$-dimensional (relativistic) CFT. With 
 NR deformation the theory however looses special conformal symmetry 
while scale 
invariance (dilatation) still exists.

\section{Discussion: CFT of cold atoms}

In this note we provided an example of a non-relativistic 
anti-de-Sitter background 
$NRadS_4\times S^6$ as the solution of 
massive type IIA supergravity. A compactification on $S^6$ 
provides
immediately the prototype NRadS background in four dimensions, which 
are generally proposed to 
be dual to a critical phenomenon involving strongly 
coupled matter (fermions)  in non-relativistic boundary theory 
 \cite{son,bala}.      

According to the  proposal \cite{son, bala} the dual field theory 
will  be a $(2+1)$-dimensional  quantum theory describing 
the non-relativistic
dynamics of strongly coupled fermions  along a {\it wire}. 
The AdS-CFT holography works well when the 
spacetime curvature and the coupling constant
are kept small in string theory,
i.e. $\Lambda$ has to be small. As we saw for the $NRadS_4$ solution, 
$\Lambda$ 
is tightly related to the mass $m_A$ of the 
$U(1)$ field and the string coupling constant $g_s$. 
We are allowed to 
have $m\to \infty$ 
only if $g_s\to 0$ keeping $m^2 g_s^{5/2}$ or $m_A^2$ fixed and small. In 
general, $m$ is 
 related to the number of D2-branes (via 4-form flux
) and/or D8-branes present in the 
background. The number of D2 and D8 branes is pretty much 
interlinked in 
massive type IIA as is evident from the solution \eqn{sol2} ($c^2=5m^2 
g_s^2$). In any situation
large $m$ would also mean large $N_c$. To recall we know that  
usual  $(2+1)$-dimensional 
(relativistic) super-Yang-Mills theory is UV complete but the theory 
flows to a strongly coupled fixed point in the
IR. A non-relativistic deformation of this theory as in this work would
then  describe nonrelativistic phenomena on a wire. 
However there remains only a
 Galilean symmetry  in IR and not the full Schr\"odinger group. We expect 
the theory will 
be having strongly coupled phases like unitary fermions on wire. It would 
be worthwhile to explore such a
non-relativistic field theory 
 as a dual of $NRadS_4\times S^6$. Incidentally
the theory will be possessing global $SO(7)$ symmetry. This global 
symmetry can reduce if,  instead of $S^6$, we select other 
compact (Einstein) 
spaces such as $CP^3$, 
$S^2 \times S^2\times S^2$, $S^2\times S^4$, or $S^3\times S^3$ 
\cite{roma}.





\end{document}